\documentclass[aps,prl,twocolumn,superscriptaddress,showpacs,preprintnumbers,amsmath,amssymb]{revtex4}

\usepackage{graphicx} 
\usepackage{dcolumn}  
\usepackage{rotating}
\usepackage{adjustbox}
\usepackage{threeparttable}
\usepackage[colorlinks,linkcolor=red,anchorcolor=green,citecolor=blue]{hyperref}

\graphicspath{{ps}}
\newcommand{\BR}{{\cal B}}
\newcommand{\EE}{e^+e^-}
\newcommand{\pp}{\pi^+\pi^-}
\newcommand {\tabincell}[2]{\begin{tabular}{@{}#1@{}}#2\end{tabular}}%

\begin{document}

\title{
 Evidence of a structure in $\bar{K}^{0} \Lambda_{c}^{+}$ consistent with a charged $\Xi_c(2930)^{+}$, and updated measurement of $\bar{B}^{0} \to \bar{K}^{0} \Lambda_{c}^{+} \bar{\Lambda}_{c}^{-}$ at Belle
}

\noaffiliation
\affiliation{University of the Basque Country UPV/EHU, 48080 Bilbao}
\affiliation{Beihang University, Beijing 100191}
\affiliation{University of Bonn, 53115 Bonn}
\affiliation{Brookhaven National Laboratory, Upton, New York 11973}
\affiliation{Budker Institute of Nuclear Physics SB RAS, Novosibirsk 630090}
\affiliation{Faculty of Mathematics and Physics, Charles University, 121 16 Prague}
\affiliation{Chonnam National University, Kwangju 660-701}
\affiliation{University of Cincinnati, Cincinnati, Ohio 45221}
\affiliation{Deutsches Elektronen--Synchrotron, 22607 Hamburg}
\affiliation{University of Florida, Gainesville, Florida 32611}
\affiliation{Key Laboratory of Nuclear Physics and Ion-beam Application (MOE) and Institute of Modern Physics, Fudan University, Shanghai 200443}
\affiliation{Justus-Liebig-Universit\"at Gie\ss{}en, 35392 Gie\ss{}en}
\affiliation{Gifu University, Gifu 501-1193}
\affiliation{II. Physikalisches Institut, Georg-August-Universit\"at G\"ottingen, 37073 G\"ottingen}
\affiliation{SOKENDAI (The Graduate University for Advanced Studies), Hayama 240-0193}
\affiliation{Gyeongsang National University, Chinju 660-701}
\affiliation{Hanyang University, Seoul 133-791}
\affiliation{University of Hawaii, Honolulu, Hawaii 96822}
\affiliation{High Energy Accelerator Research Organization (KEK), Tsukuba 305-0801}
\affiliation{J-PARC Branch, KEK Theory Center, High Energy Accelerator Research Organization (KEK), Tsukuba 305-0801}
\affiliation{Forschungszentrum J\"{u}lich, 52425 J\"{u}lich}
\affiliation{IKERBASQUE, Basque Foundation for Science, 48013 Bilbao}
\affiliation{Indian Institute of Science Education and Research Mohali, SAS Nagar, 140306}
\affiliation{Indian Institute of Technology Bhubaneswar, Satya Nagar 751007}
\affiliation{Indian Institute of Technology Guwahati, Assam 781039}
\affiliation{Indian Institute of Technology Hyderabad, Telangana 502285}
\affiliation{Indian Institute of Technology Madras, Chennai 600036}
\affiliation{Indiana University, Bloomington, Indiana 47408}
\affiliation{Institute of High Energy Physics, Chinese Academy of Sciences, Beijing 100049}
\affiliation{Institute of High Energy Physics, Vienna 1050}
\affiliation{INFN - Sezione di Napoli, 80126 Napoli}
\affiliation{INFN - Sezione di Torino, 10125 Torino}
\affiliation{Advanced Science Research Center, Japan Atomic Energy Agency, Naka 319-1195}
\affiliation{J. Stefan Institute, 1000 Ljubljana}
\affiliation{Kanagawa University, Yokohama 221-8686}
\affiliation{Institut f\"ur Experimentelle Kernphysik, Karlsruher Institut f\"ur Technologie, 76131 Karlsruhe}
\affiliation{King Abdulaziz City for Science and Technology, Riyadh 11442}
\affiliation{Department of Physics, Faculty of Science, King Abdulaziz University, Jeddah 21589}
\affiliation{Korea Institute of Science and Technology Information, Daejeon 305-806}
\affiliation{Korea University, Seoul 136-713}
\affiliation{Kyungpook National University, Daegu 702-701}
\affiliation{LAL, Univ. Paris-Sud, CNRS/IN2P3, Universit\'{e} Paris-Saclay, Orsay}
\affiliation{\'Ecole Polytechnique F\'ed\'erale de Lausanne (EPFL), Lausanne 1015}
\affiliation{P.N. Lebedev Physical Institute of the Russian Academy of Sciences, Moscow 119991}
\affiliation{Faculty of Mathematics and Physics, University of Ljubljana, 1000 Ljubljana}
\affiliation{Ludwig Maximilians University, 80539 Munich}
\affiliation{Luther College, Decorah, Iowa 52101}
\affiliation{University of Malaya, 50603 Kuala Lumpur}
\affiliation{University of Maribor, 2000 Maribor}
\affiliation{Max-Planck-Institut f\"ur Physik, 80805 M\"unchen}
\affiliation{School of Physics, University of Melbourne, Victoria 3010}
\affiliation{University of Mississippi, University, Mississippi 38677}
\affiliation{Moscow Physical Engineering Institute, Moscow 115409}
\affiliation{Moscow Institute of Physics and Technology, Moscow Region 141700}
\affiliation{Graduate School of Science, Nagoya University, Nagoya 464-8602}
\affiliation{Kobayashi-Maskawa Institute, Nagoya University, Nagoya 464-8602}
\affiliation{Universit\`{a} di Napoli Federico II, 80055 Napoli}
\affiliation{National Central University, Chung-li 32054}
\affiliation{National United University, Miao Li 36003}
\affiliation{Department of Physics, National Taiwan University, Taipei 10617}
\affiliation{H. Niewodniczanski Institute of Nuclear Physics, Krakow 31-342}
\affiliation{Nippon Dental University, Niigata 951-8580}
\affiliation{Niigata University, Niigata 950-2181}
\affiliation{Novosibirsk State University, Novosibirsk 630090}
\affiliation{Osaka City University, Osaka 558-8585}
\affiliation{Pacific Northwest National Laboratory, Richland, Washington 99352}
\affiliation{Panjab University, Chandigarh 160014}
\affiliation{Peking University, Beijing 100871}
\affiliation{University of Pittsburgh, Pittsburgh, Pennsylvania 15260}
\affiliation{Theoretical Research Division, Nishina Center, RIKEN, Saitama 351-0198}
\affiliation{University of Science and Technology of China, Hefei 230026}
\affiliation{Showa Pharmaceutical University, Tokyo 194-8543}
\affiliation{Soongsil University, Seoul 156-743}
\affiliation{Stefan Meyer Institute for Subatomic Physics, Vienna 1090}
\affiliation{Sungkyunkwan University, Suwon 440-746}
\affiliation{School of Physics, University of Sydney, New South Wales 2006}
\affiliation{Department of Physics, Faculty of Science, University of Tabuk, Tabuk 71451}
\affiliation{Tata Institute of Fundamental Research, Mumbai 400005}
\affiliation{Department of Physics, Technische Universit\"at M\"unchen, 85748 Garching}
\affiliation{Department of Physics, Tohoku University, Sendai 980-8578}
\affiliation{Department of Physics, University of Tokyo, Tokyo 113-0033}
\affiliation{Tokyo Institute of Technology, Tokyo 152-8550}
\affiliation{Tokyo Metropolitan University, Tokyo 192-0397}
\affiliation{Virginia Polytechnic Institute and State University, Blacksburg, Virginia 24061}
\affiliation{Wayne State University, Detroit, Michigan 48202}
\affiliation{Yamagata University, Yamagata 990-8560}
\affiliation{Yonsei University, Seoul 120-749}
\author{Y.~B.~Li}\affiliation{Peking University, Beijing 100871} 
\author{C.~P.~Shen}\affiliation{Beihang University, Beijing 100191} 
\author{I.~Adachi}\affiliation{High Energy Accelerator Research Organization (KEK), Tsukuba 305-0801}\affiliation{SOKENDAI (The Graduate University for Advanced Studies), Hayama 240-0193} 
\author{H.~Aihara}\affiliation{Department of Physics, University of Tokyo, Tokyo 113-0033} 
\author{S.~Al~Said}\affiliation{Department of Physics, Faculty of Science, University of Tabuk, Tabuk 71451}\affiliation{Department of Physics, Faculty of Science, King Abdulaziz University, Jeddah 21589} 
\author{D.~M.~Asner}\affiliation{Brookhaven National Laboratory, Upton, New York 11973} 
\author{T.~Aushev}\affiliation{Moscow Institute of Physics and Technology, Moscow Region 141700} 
\author{R.~Ayad}\affiliation{Department of Physics, Faculty of Science, University of Tabuk, Tabuk 71451} 
\author{V.~Babu}\affiliation{Tata Institute of Fundamental Research, Mumbai 400005} 
\author{I.~Badhrees}\affiliation{Department of Physics, Faculty of Science, University of Tabuk, Tabuk 71451}\affiliation{King Abdulaziz City for Science and Technology, Riyadh 11442} 
% ...
\collaboration{The Belle Collaboration}

\begin{abstract}

We report evidence for the charged charmed-strange baryon $\Xi_{c}(2930)^+$ with a signal significance of 3.9$\sigma$ with systematic errors included.
The charged $\Xi_{c}(2930)^+$ is found in its decay to $K_{S}^{0} \Lambda_{c}^+$ in the substructure of $\bar{B}^{0} \to K^{0}_{S} \Lambda_{c}^{+} \bar{\Lambda}_{c}^{-}$ decays. The measured mass and width are $[2942.3 \pm 4.4 (\rm stat.) \pm 1.5(\rm syst.)]$~MeV/$c^{2}$ and $[14.8 \pm 8.8(\rm stat.) \pm 2.5(\rm syst.)]$~MeV, respectively, and the product branching fraction is $\BR(\bar{B}^{0} \to \Xi_c(2930)^{+} \bar{\Lambda}_{c}^{-}) \BR(\Xi_c(2930)^{+}\to \bar{K}^{0} \Lambda_{c}^{+})=[2.37 \pm 0.51 (\rm stat.)\pm 0.31(\rm syst.)]\times 10^{-4}$. We also measure $\BR(\bar{B}^{0} \to \bar{K}^{0} \Lambda_{c}^{+} \bar{\Lambda}_{c}^{-}) = [3.99 \pm 0.76(\rm stat.) \pm 0.51(\rm syst.)] \times 10^{-4}$ with greater precision than previous experiments, and present the results of a search for the charmonium-like state $Y(4660)$ and its spin partner, $Y_{\eta}$, in the $\Lambda_{c}^{+}\bar{\Lambda}_{c}^{-}$
invariant mass spectrum. No clear signals of the $Y(4660)$ or $Y_{\eta}$ are observed and the 90\% credibility level (C.L.) upper limits on their production rates are determined. These measurements are obtained from a sample of $(772\pm11)\times 10^{6} B\bar{B}$ pairs collected at the $\Upsilon(4S)$ resonance by the Belle detector at the KEKB asymmetric energy electron-positron collider.
\end{abstract}


\maketitle

\tighten

{\renewcommand{\thefootnote}{\fnsymbol{footnote}}}
\setcounter{footnote}{0}

The study of the excited states of charmed and bottom baryons is important as they offer an excellent laboratory for testing the heavy-quark symmetry of the $c$ and $b$ quarks and the chiral symmetry of the light quarks. At present, the particle data group (PDG) lists ten charmed-strange baryons~\cite{PDG}. Among these, $\Xi_c(2930)$ and $\Xi_c(3123)$ are relatively less established and the evidence for them is poor~\cite{PDG}. For most of these excited $\Xi_c$ states the spin and parity ($J^P$) have not been determined by experiments due to limited statistics. 

Theoretically, the mass spectrum of excited charmed baryons has been computed in many models, including quark potential models~\cite{Majethiya,Migura, Garcilaz, Ebert1, Ebert2}, the relativistic flux tube model~\cite{Chen1, Chen2}, the coupled channel model~\cite{Romanets}, the  Quantum Chromodynamics (QCD) sum rule~\cite{Wang1, Wang2, Ye, sundu, Chen3}, Regge phenomenology~\cite{Guo}, the constituent quark model~\cite{kai,bing},  and lattice QCD~\cite{Padmanath1, Padmanath2}. The strong decays of excited $\Xi_c$ baryons have also been studied in many models~\cite{Chen4, Cheng1, Cheng2, Chen5, Liu, Zhao, Chen6}. In these models, some possible $J^P$ assignments of these excited $\Xi_c$ have been performed. While many new excited charmed baryons have been discovered in experiments in recent years, and there has been dedicated theoretical work devoted to study the nature of charmed baryon such as the baryon internal structure and quark configuration, further cooperative efforts are needed from both experimentalists and theorists to make progress in this area.

Very recently, Belle reported the first observation of the $\Xi_{c}(2930)^0$ charmed-strange baryon with a significance greater than 5$\sigma$
from a study of the substructure of $B^{-} \to K^{-} \Lambda_{c}^{+} \bar{\Lambda}_{c}^{-}$ decays~\cite{myana}. The measured mass and width of the $\Xi_{c}(2930)^0$ were found to be $[2928.9 \pm 3.0(\rm stat.)^{+0.9}_{-12.0}(\rm syst.)]$~MeV/$c^{2}$ and $[19.5 \pm 8.4(\rm stat.)^{+5.9}_{-7.9}(\rm syst.)]$~MeV, respectively. As the isospin of the $\Xi_{c}$ state is always $\frac{1}{2}$ and the neutral $\Xi_{c}(2930)^0$ has been found, it is natural to search for the charged $\Xi_{c}(2930)^+$ state in the substructure in $\bar{B}^{0} \to \bar{K}^{0} \Lambda_{c}^{+}\bar{\Lambda}_{c}^{-}$ decays.

BaBar and Belle have previously studied $\bar{B}^{0} \to \bar{K}^{0} \Lambda_{c}^{+} \bar{\Lambda}_{c}^{-}$ decays using data samples of $230 \times 10^{6}$ and $386\times 10^{6}~B\bar{B}$ pairs, and found signals of $1.4\sigma$ and $6.6 \sigma$ significances, respectively~\cite{barbar,belleold}. Neither experiment searched for possible intermediate states such as the $K_{S}^{0} \Lambda_c$ system. The full Belle data sample of $(772\pm11)\times 10^{6} B\bar{B}$ pairs permits an improved study of $\bar{B}^{0} \to \bar{K}^{0} \Lambda_{c}^{+} \bar{\Lambda}_{c}^{-}$ and a search for the charged $\Xi_{c}(2930)^+$ in the decay mode $\bar{K}^{0}\Lambda_{c}^+$.

The $\Lambda_{c}^{+} \bar{\Lambda}_{c}^{-}$ system is interesting because (1)
Belle has observed the $Y(4630)$ in the initial state radiation (ISR) process $\EE \to \gamma_{\rm ISR} \Lambda_{c}^{+} \bar{\Lambda}_{c}^{-}$ and measured a mass and width of $[4634^{+8}_{-7}({\rm stat.})^{+5}_{-8}({\rm syst.})]$ MeV/$c^2$ and $[92^{+40}_{-24}({\rm stat.}) ^{+10}_{-21}({\rm syst.})]$ MeV, respectively~\cite{4630}; (2) Belle has also observed the $Y(4660)$ in $e^+e^- \to \gamma_{\rm ISR}\pp\psi'$ with a measured mass and width of $[4652\pm10({\rm stat.})\pm 8({\rm syst.})]$  MeV/$c^2$ and $[68\pm 11({\rm stat.})\pm 1({\rm syst.})]$ MeV, respectively~\cite{Belle4660isr-old,Belle4660isr-new}. As the masses and widths of the $Y(4630)$ and $Y(4660)$ are close to each other,  many theoretical explanations assume they are the same state~\cite{2Ysame1,2Ysame2,4616b}. In Refs.~\cite{thero4600,4616a}, the authors predicted a $Y(4660)$ spin partner---a $f_{0}(980)\eta_{c}(2S)$ bound state denoted by the $Y_{\eta}$---with a mass and width of $(4613 \pm 4)$ MeV/$c^2$ and around 30 MeV, respectively, with the assumption that the $Y(4660)$ is an $f_{0}(980)\psi'$ bound state~\cite{4616a,4616b}. Belle has searched for these states in the substructure of $B^{-} \to K^{-} \Lambda_{c}^{+}\bar{\Lambda}_{c}^{-}$ decays, and no clear signals were observed~\cite{myana}. The corresponding $B^0$ decay mode can also be used to study the $\Lambda_{c}^{+} \bar{\Lambda}_{c}^{-}$ invariant mass.

In this letter, we report an updated measurement of $\bar{B}^{0} \to \bar{K}^{0} \Lambda_{c}^{+} \bar{\Lambda}_{c}^{-}$ and a search for the charged $\Xi_{c}(2930)^+\to \bar{K}^{0} \Lambda_{c}^{+}$ state with a statistical significance of 4.1$\sigma$~\cite{charge-conjugate}. This analysis is based on the full data sample collected at the $\Upsilon(4S)$ resonance by the Belle detector~\cite{Belle} at the KEKB asymmetric energy electron-positron collider~\cite{KEKB}.

The Belle detector is a large solid angle magnetic spectrometer
that consists of a silicon vertex detector, a 50-layer
central drift chamber (CDC), an array of aerogel threshold
Cherenkov counters (ACC), a barrel-like arrangement of
time-of-flight scintillation counters (TOF), and an
electromagnetic calorimeter comprised of CsI(Tl) crystals located
inside a superconducting solenoid coil that provides a $1.5~\hbox{T}$
magnetic field. An iron flux-return yoke located outside the coil is
instrumented to detect $K^{0}_{L}$ mesons and to identify muons.
A detailed description of the Belle detector
can be found in Ref.~\cite{Belle}.
Simulated signal events with $B$ meson decays are generated using {\sc EvtGen}~\cite{evtgen}, while the inclusive decays are generated via {\sc PYTHIA}~\cite{pythia}. These events are  processed by a detector simulation based on {\sc GEANT3}~\cite{geant3}. Inclusive Monte Carlo (MC) samples of $\Upsilon(4S)\to B \bar{B}$ ($B=B^+$ or $B^0$) and $e^+e^- \to q \bar{q}$ ($q=u,~d,~s,~c$) events at $\sqrt{s}=10.58$ GeV are used to check the backgrounds, corresponding to more than 5 times the integrated luminosity of the data.


In our analysis of $\bar{B}^{0} \to \bar{K}^0 \Lambda_c^+ \bar{\Lambda}_{c}^{-}$, $\bar{K}^{0}$ is reconstructed via its decay $K_{S}^{0}\to \pi^+\pi^-$,
and $\Lambda_c^+$ candidates are reconstructed in the $\Lambda_c^+\to p K^- \pi^+$, $pK_{S}^{0}$, and $\Lambda \pi^{+}(\to p \pi^{-}\pi^+)$ decay channels.
Then a $\Lambda_c^+$ and $\bar{\Lambda}_{c}^{-}$ are combined to reconstruct a $B$ candidate, with at least one required to have been reconstructed via the $p K^- \pi^+$ or $\bar{p} K^+ \pi^-$ decay process.

For well reconstructed charged tracks, except for those from $\Lambda \to p \pi^{-}$ and $K_{S}^{0} \to \pi^{+} \pi^{-}$ decays, the impact parameters perpendicular to and along the beam direction with respect to the nominal interaction point are required to be less than 0.5 cm and 4 cm, respectively, and the transverse momentum in the laboratory frame is required to be larger than 0.1~GeV/$c$. The information from different detector subsystems including specific ionization in the CDC, time measurements in the TOF and response of the ACC is combined to form the likelihood ${\mathcal L}_i$ of the track for particle species $i$, where $i=\pi$,~$K$ or $p$~\cite{pid}. Except for the charged tracks from $\Lambda \to p \pi^{-}$ and $K_{S}^{0} \to \pi^{+} \pi^{-}$ decays, tracks with a likelihood ratio $\mathcal{R}_K^{\pi} = \mathcal{L}_K/(\mathcal{L}_K + \mathcal{L}_\pi)> 0.6$ are identified as kaons, while tracks with $\mathcal{R}_K^{\pi}<0.4$ are treated as pions. The kaon (pion) identification efficiency is about 94\% (97\%), while 5\% (3\%) of the kaons (pions) are misidentified as pions (kaons) with the selection criteria above.  For proton identification, a track with $\mathcal{R}^\pi_{p/\bar{p}} = \mathcal{L}_{p/\bar{p}}/(\mathcal{L}_{p/\bar{p}}+\mathcal{L}_\pi) > 0.6$ and $\mathcal{R}^K_{p/\bar{p}} = \mathcal{L}_{p/\bar{p}}/(\mathcal{L}_{p/\bar{p}}+\mathcal{L}_K) > 0.6$ is identified as a proton/anti-proton with an efficiency of about 98\%; less than $1\%$ of the pions/kaons are misidentified as protons/anti-protons.

The $K_{S}^{0}$ candidates are reconstructed from pairs of oppositely-charged tracks which are treated as pions, and identified by a multivariate analysis with a neural network~\cite{NN} based on two sets of input variables~\cite{NN-input}. Candidate $\Lambda$ baryons are reconstructed in the decay $\Lambda \to p \pi^-$ and selected if the $p \pi^-$ invariant mass is within 5 MeV/$c^2$ (5$\sigma$) of the $\Lambda$ nominal mass~\cite{PDG}.

A vertex fit to the $B$ candidates is performed and the candidate with the minimum $\chi^{2}_{\rm vertex}/n.d.f.$ from the vertex fit is selected as the signal $B$ candidate if there is more than one $B$ candidates in an event, where $n.d.f.$ is the number of freedom of the vertex fit. Then $\chi^{2}_{\rm vertex}/n.d.f. <15$ is required, which has a selection efficiency above 96\%. As the continuum background level is very low, further continuum suppression is not necessary.

\begin{figure}[htbp]
\begin{center}
\includegraphics[width=4.17cm]{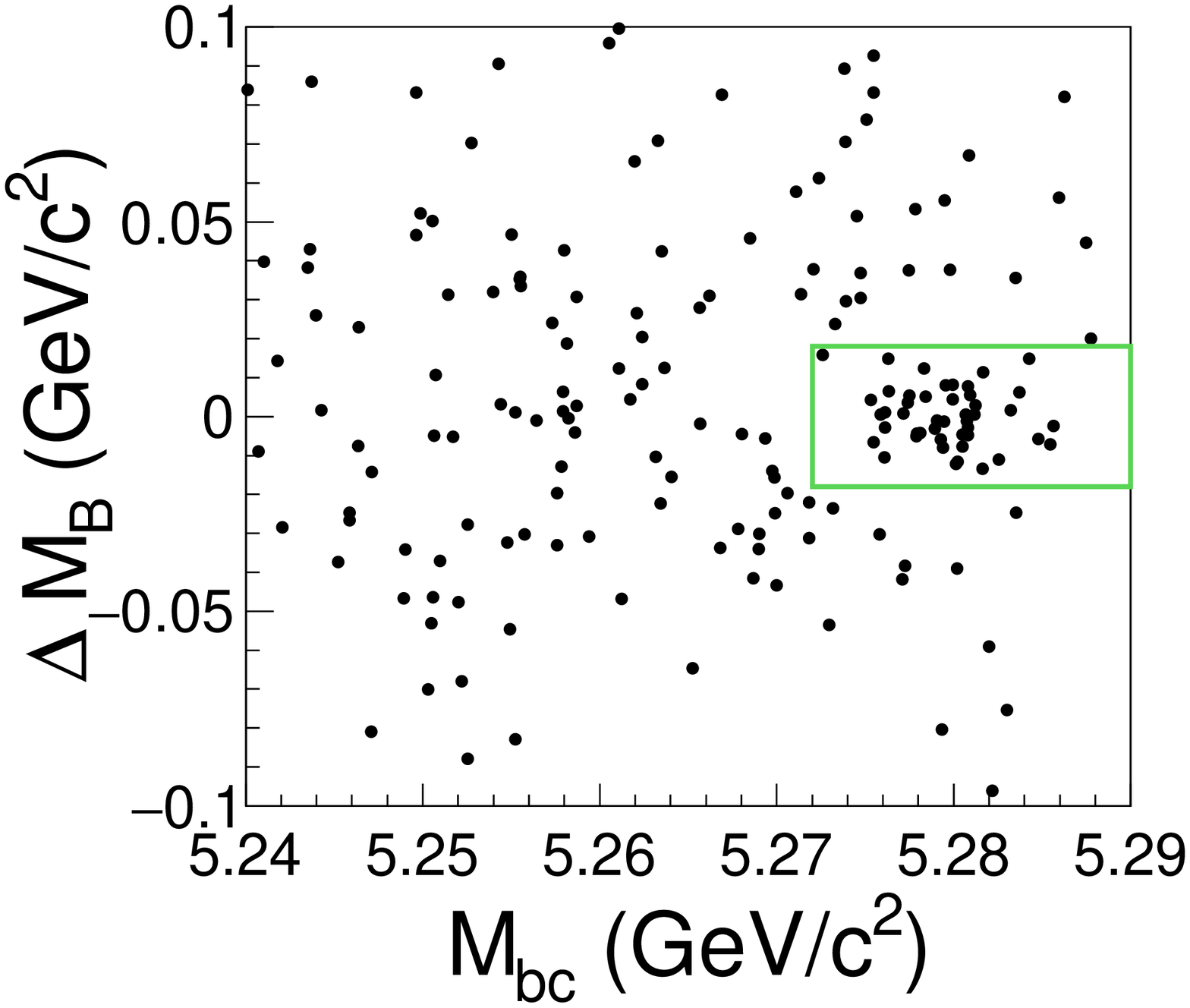}
\includegraphics[width=4.36cm]{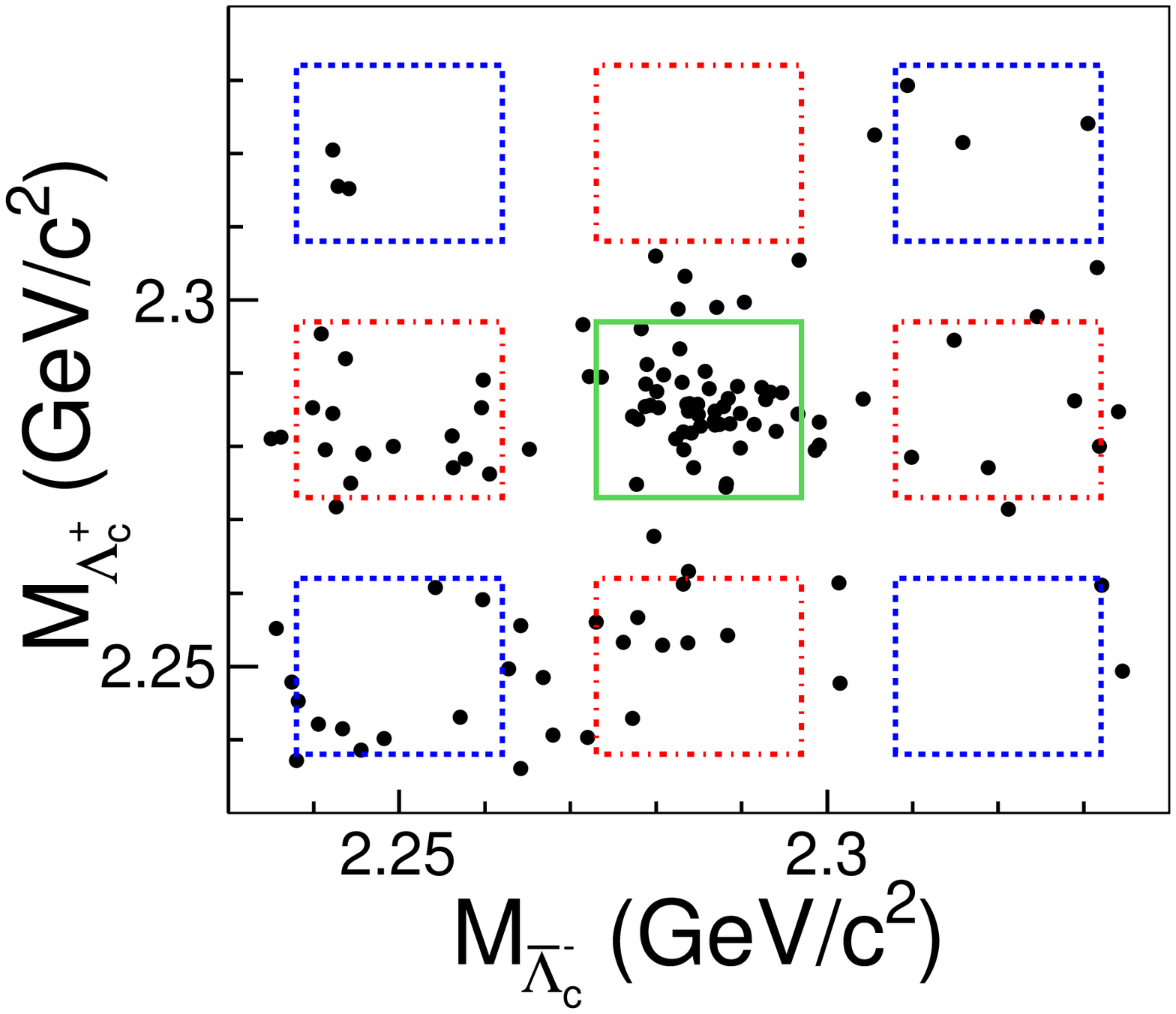}
\caption{\label{comlbdac_bsignal}
Signal-enhanced distributions of $\Delta M_{B}$ versus $M_{\rm bc}$ (left panel) and of $M(\bar{\Lambda}_{c}^{-})$ versus $M(\Lambda_{c}^{+})$ (right panel) from the selected $\bar{B}^{0} \to \bar{K}^{0} \Lambda_{c}^{+} \bar{\Lambda}_{c}^{-}$ candidates, summing over all three reconstructed $\Lambda_{c}$ decay modes. Each panel shows the events falling in the solid green signal region of the other panel. The dashed red and blue boxes in the left panel show the defined $\Lambda_c$ sideband regions described in the text,
which are  used for the estimation of the non-$\Lambda_c$ background.}
\end{center}
\end{figure}

The $B$ candidates are identified using the beam-energy constrained mass $M_{\rm bc}$ and the mass difference $\Delta M_{B}$. The beam-energy constrained mass is defined as $M_{\rm bc} \equiv \sqrt{E_{\rm beam}^{2}/c^2 - (\sum \vec{p}_{i})^2}/c$, where $E_{\rm beam}$ is the beam energy and $\vec{p}_{i}$ are the three-momenta of the $B$-meson decay products, all defined in the center-of-mass system (CMS) of the $\EE$ collision. The mass difference is defined as $\Delta M_{B} \equiv M_{B} - m_{B}$, where $M_{B}$ is the invariant mass of the $B$ candidate and $m_{B}$ is the nominal $B$-meson mass~\cite{PDG}. The $B$ signal region is defined as $|\Delta M_B| < 0.018~{\rm GeV}/c^2$ and $M_{\rm bc} > 5.272\,{\rm GeV}/c^2$ ($\sim 2.5 \sigma$) which is shown as the central box in the distribution of $\Delta M_B$ versus $M_{\rm bc}$ in Fig~\ref{comlbdac_bsignal}.
	

The scatter plot of $M_{\bar{\Lambda}_{c}^{-}}$ versus $M_{\Lambda_{c}^{+}}$ is shown in the right panel of Fig.~\ref{comlbdac_bsignal} for the selected $\bar{B}^{0} \to K^{0}_{S} \Lambda_{c}^{+} \bar{\Lambda}_{c}^{-}$ data candidates in the $B$ signal region, and clear $\Lambda_c^+$ and $\bar{\Lambda}_c^-$ signals are observed. According to the signal MC simulation, the mass resolution of $\Lambda_{c}$ candidates is almost independent of the $\Lambda_{c}$ decay mode. The $\Lambda_{c}$ signal region is defined as $|M_{\Lambda_{c}} - m_{\Lambda_{c}} |< 12$ MeV/$c^{2}$ ($\sim 2.5\sigma$) for all $\Lambda_{c}$ decay modes illustrated by the central green box in the Fig.~\ref{comlbdac_bsignal} (right panel), where $m_{\Lambda_{c}}$ is the nominal mass of the $\Lambda_{c}$ baryon~\cite{PDG}. To estimate the non-$\Lambda_c$ backgrounds, we define the $\Lambda_{c}^{+}$ and $\bar{\Lambda}_{c}^{-}$ mass sidebands as half of the total number of events in the four
sideband regions next to the signal region minus one quarter of the total number of events in the four
sideband regions in the corners as shown in Fig.~\ref{comlbdac_bsignal} (right panel).

To extract the $\bar{B}^{0} \to K^{0}_{S} \Lambda_{c}^{+} \bar{\Lambda}^{-}_{c}$ signal yields, we perform an unbinned two-dimensional (2D) simultaneous extended maximum likelihood fit to the $\Delta M_{B}$ versus $M_{\rm bc}$ distributions for the three reconstructed $\Lambda_{c}$ decay modes. A Gaussian function for the signal shape plus an ARGUS function~\cite{argus} for the background are used to fit the $M_{\rm bc}$ distribution, and the sum of a double-Gaussian function for the signal plus a first-order polynomial for the background are used to fit the $\Delta M_{B}$ distribution. Due to limited statistics, all the parameters of the Gaussian functions are fixed to the values from the fits to the individual MC signal distributions, and the relative signal yields among the three final states are fixed according to the relative branching fraction between the final states and the detection acceptance and efficiency of the intermediate states.

\begin{figure}[htbp]
\begin{center}
\includegraphics[width=4.2cm,height=3.3cm]{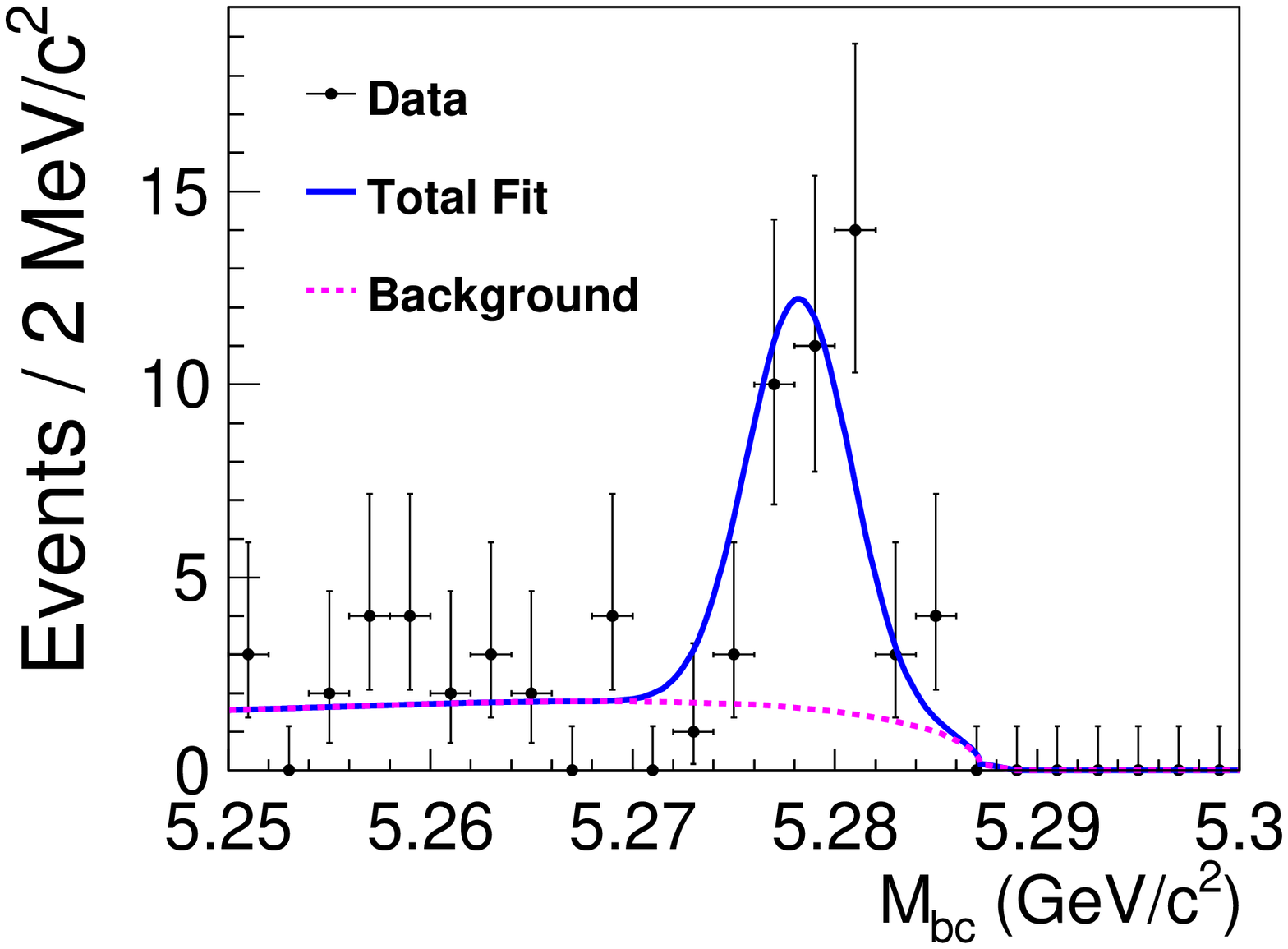}
\includegraphics[width=4.2cm,height=3.3cm]{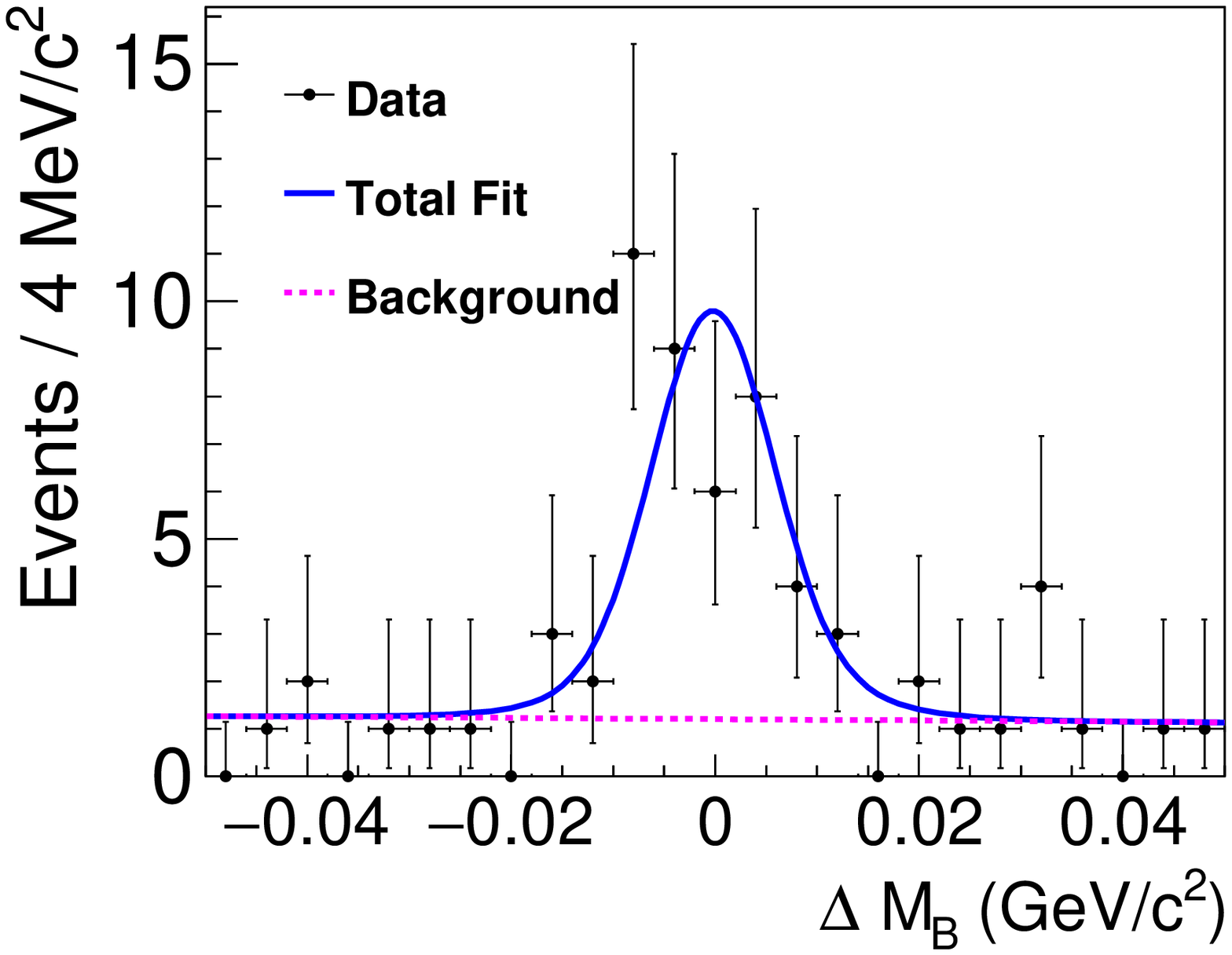}
\put(-160,70){\bf (a)}
\put(-40,70){\bf (b)}
\caption{\label{MvsEcom} The $\Lambda_c$-signal-enhanced distributions of (a) $M_{\rm bc}$ in the $\Delta M_{B}$ signal region and (b) $\Delta M_{B}$ in the $M_{\rm bc}$ signal region for $B^{0} \to K^{0}_{S} \Lambda_{c}^{+} \bar{\Lambda}^{-}_{c}$, combining three exclusive final states. The dots with error bars are data, the solid blue curves are the best-fit projections to the distributions, and the dashed magenta lines are the fitted backgrounds.}
\end{center}
\end{figure}

The projections of $M_{\rm bc}$ and $\Delta M_{B}$ summed over the three reconstructed $\Lambda_{c}$ decay modes
in $\Lambda_{c}$ signal region, together with the fitted results, are shown in Fig.~\ref{MvsEcom}.
There are $34.9 \pm 6.6$ signal events with a statistical signal significance above 8.3$\sigma$,
and from which we extract the branching fraction of $\BR (\bar{B}^{0} \to \bar{K}^{0}
\Lambda_{c}^{+} \bar{\Lambda}_{c}^{-})=[3.99 \pm 0.76({\rm stat.})] \times 10^{-4}$.

\begin{figure}[htbp]
\begin{center}
\includegraphics[width=6cm]{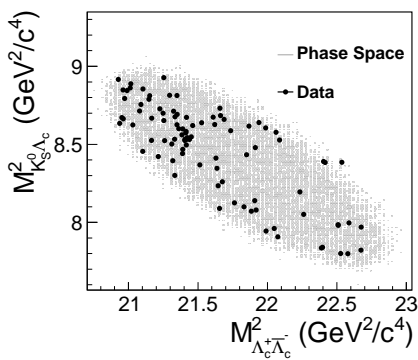}
\caption{\label{sigScat} Dalitz distribution of reconstructed $\bar{B}^{0} \to
K_S^0 \Lambda_{c}^{+} \bar{\Lambda}_{c}^{-}$ candidates in the $B$
signal region. The black dots are data; the shaded region is
the MC simulated phase-space distribution.}
\end{center}
\end{figure}

To check the intermediate states, mass constraint fits of $K_S^{0}$, $\bar{\Lambda}_{c}^{-}$, and $\bar{B}^{0}$
are applied to the selected candidates in the signal regions to improve the mass resolutions, while for the
above defined sidebands no mass constraint fits are applied.
After applying all selection criteria above, Dalitz distribution of the $M^{2}_{K^{0}_{S} \Lambda_{c}}$ versus $ M^{2}_{\Lambda_{c}^{+} \bar{\Lambda}_{c}^{-}}$ is shown in Fig.~\ref{sigScat} with a flat 2D efficiency distribution. Here, $M^{2}_{K^{0}_{S} \Lambda_{c}}$ is the sum of $M^{2}_{K^{0}_{S} \Lambda_{c}^{+}}$ and $M^{2}_{K^{0}_{S} \bar{\Lambda}_{c}^{-}}$. An enhancement can be seen in the horizontal band corresponding to $M(K^{0}_{S} \Lambda_{c}) \sim 2.93$ GeV/$c^2$, while no signal band is apparent in the $M(\Lambda_{c}^{+} \bar{\Lambda}_c^-)$ vertical direction.

\begin{figure}[htbp]
		\includegraphics[width=6.4cm]{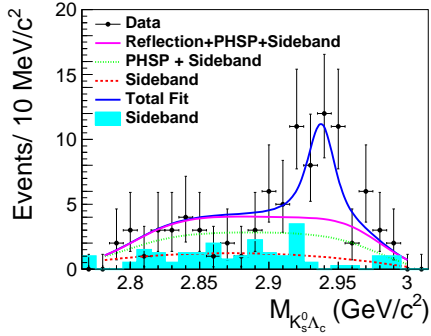}
		\vspace{-0.3cm}
		\caption{ \label{lkll}
			The $M_{K^{0}_{S}\Lambda_{c}^+}$ distribution of the selected data candidates, with fit results superimposed. Dots with error bars are the data, the solid blue line is the best fit,the solid magenta line is the total non-charged-$\Xi_{c}(2930)$ backgrounds including the fitted phase space, the reflection of the charged $\bar{\Xi}_{c}(2930)^-$, and the fitted sideband shape, the dotted green line is the fitted phase space and sideband shape, the dotted red line is the fitted sideband shape, the shaded cyan histogram is from the normalized $\Lambda_{c}^{+}$ and $\bar{\Lambda}_{c}^{-}$ mass sidebands.}

\end{figure}

The sum of the projections of $M_{K^{0}_{S} \Lambda_{c}^{+}}$ and $M_{K^{0}_{S} \bar{\Lambda}_{c}^{-}}$ mass spectra, denoted $M_{K^{0}_{S} \Lambda_{c}}$, is shown in Fig.~\ref{lkll}. The shaded histogram is from the normalized $\Lambda_{c}^{+}$ and $\bar{\Lambda}_{c}^{-}$ mass sidebands, which is consistent with the contributions
from normalized $\EE \to q\bar{q}$ and $\Upsilon(4S) \to B
\bar{B}$ generic MC samples. Therefore, the estimate from
the normalized $\Lambda_{c}^{+}$ and $\bar{\Lambda}_{c}^{-}$ mass sidebands
is taken to represent the total background, neglecting
the small possible contribution of background with real $\Lambda_{c}^{+}$ and $\bar{\Lambda}_{c}^{-}$.

A clear charged $\Xi_{c}(2930)^+$ signal is found. No structure is seen in the $\Lambda_{c}^{+}$ and $\bar{\Lambda}_{c}^{-}$ mass sidebands.

An unbinned simultaneous extended maximum likelihood fit is performed to the $K^{0}_{S} \Lambda_{c}^+$ invariant mass spectra for the total selected signal candidates and the $\Lambda_{c}^{+}$ and $\bar{\Lambda}_{c}^{-}$ mass sidebands. The following components are included in the fit to
the $K^{0}_{S} \Lambda_{c}^+$ mass distribution for the total selected signal candidates:
a constant width relativistic Breit-Wigner (RBW) function ($\frac{1}{M^{2}_{\Xi^{+}_{c}(2930)} - M^{2}_{K_{S}^{0}\Lambda_{c}} - i M_{\Xi^{+}_{c}(2930)} \Gamma_{\Xi^{+}_{c}(2930)}}$) convolved with a Gaussian resolution function with the phase space factor and efficiency curve included (the width of the Gaussian function being fixed to 5.36 MeV/$c^2$ from the signal MC simulation) is taken as the charged $\Xi_{c}(2930)^+$ signal shape;
a broader structure obtained by MC simulation is used to represent
the reflection of the charged $\bar{\Xi}_{c}(2930)^-$;
direct three-body $\bar{B}^{0} \to K^{0}_{S} \Lambda_{c}^{+} \bar{\Lambda}_{c}^{-}$ decays are modeled
by the MC-simulated shape distributed uniformly in phase space;
a second-order polynomial is used to represent the
$\Lambda_{c}^{+}$ and $\bar{\Lambda}_{c}^{-}$ mass-sideband
distribution, which is normalized to represent the
total background events in the fit.
In the above fit, the signal yields of the charged $\Xi_{c}(2930)^+$
and the corresponding reflection are constrained to be the same.

The fit results are shown in Fig.~\ref{lkll}, where the solid blue line is the best fit, and the solid magenta line is the total non-charged-$\Xi_{c}(2930)$ backgrounds including the fitted phase space, the reflection of the  charged $\bar{\Xi}_{c}(2930)^{-}$, and the fitted sideband shape. The yields of the charged $\Xi_{c}(2930)^+$ signal and the phase-space contribution are $N_{\Xi_{c}(2930)^+} = 21.2 \pm 4.6$ and $N_{phsp} = 18.3 \pm 4.6$. The fitted mass and width are $M_{\Xi_{c}(2930)^{+}} = [2942.3\pm 4.4 (\rm stat.)]$ MeV/$c^{2}$ and $\Gamma_{\Xi_{c}(2930)^{+}} = [14.8 \pm 8.8 (\rm stat.)] $ MeV, respectively, where the correction of 2.8 MeV/$c^2$ has been applied on the charged $\Xi_{c}(2930)^+$ mass, determined using the input and output mass difference in the MC simulation.
The statistical significance of the charged $\Xi_{c}(2930)^+$ signal is $4.1\sigma$, calculated from the difference of the logarithmic likelihoods~\cite{significance}, $-2\ln(\mathcal{L}_{0}/\mathcal{L}_{\rm max}) = 23.1$, where $\mathcal{L}_{0}$ and $\mathcal{L}_{\rm max}$ are the maximized likelihoods without and with a signal component, respectively, taking into account the difference in the number of degrees of freedom ($\Delta$ndf = 3). The signal significance is 3.9$\sigma$ when convolving the likelihood profile with a Gaussian function of width equals the total systematic uncertainty from detection efficiency, fitting procedure, intermediate states' branching fractions. Alternative fits to the $K^{0}_{S}\Lambda_{c}$ mass spectra are performed: (a) using a first-order or third-order polynomial for background shape; (b) changing the charged $\Xi_{c}(2930)^+$ mass resolution by 10\%; and (c) using an energy-dependent RBW function for the charged $\Xi_{c}(2930)^+$ signal shape. The charged $\Xi_{c}(2930)^+$ signal significance is larger than $3.5\sigma$ in all cases.

The product branching fraction of $\BR(\bar{B}^{0} \to \Xi_{c}(2930)^{+} \bar{\Lambda}_{c}^{-})$$\BR(\Xi_{c}(2930)^{+} \to \bar{K}^{0} \Lambda_{c}^{+})$ is $[2.37 \pm 0.51(\rm stat.)]\times 10^{-4}$ calculated as $\frac{N_{\rm \Xi_{c}(2930)^{+}}}{\varepsilon^{\Xi_c(2930)^{+}}_{\rm all} N_{B^{0}\bar{B}^{0}}\BR(\Lambda_{c}^{+} \to pK^{-} \pi^{+})^{2}}$, where $N_{\rm \Xi_{c}(2930)^{+}}$ is the fitted charged $\Xi_{c}(2930)^+$ signal yield; $N_{B^{0}\bar{B}^{0}}=N_{\Upsilon(4S)}\BR(\Upsilon(4S)\to B^0\bar{B}^0)$ ($N_{\Upsilon(4S)}$ is the number of accumulated $\Upsilon(4S)$ events and $\BR(\Upsilon(4S)\to B^{0}\bar{B}^{0})=0.486 \pm 0.006$~\cite{PDG}); $\BR(\Lambda_c^+\to p K^- \pi^+)=(6.23 \pm 0.33)\%$ is the world-average branching fraction for $\Lambda_c^+\to p K^- \pi^+$~\cite{PDG}; $\varepsilon^{\Xi_c(2930)^{+}}_{\rm all}= \Sigma \varepsilon^{\Xi_c(2930)^{+}}_{i} \Gamma_{i}/\Gamma(p K^{-} \pi^{+})$ ($i$ is the $\Lambda_c$ decay-mode index, $\varepsilon_{i}^{\Xi_{c}(2930)^{+}}$ is the detection efficiency by fitting the $M_{K_{S}^{0} \Lambda_{c}^+}$ spectrum from signal MC with a charged $\Xi_{c}(2930)^+$ intermediate state, and $\Gamma_{i}$ is the partial decay width of $\Lambda_{c}^{+} \to p K^{-} \pi^{+},~pK_{S}^0$, and $\Lambda \pi^{-}$~\cite{PDG}). Here, $\BR(K_{S}^0 \to \pp) $ or $\BR(\Lambda \to p \pi^{-})$ is included in $\Gamma_{i}$ for the final states with a $K_{S}^0$ or a $\Lambda$.

\begin{figure}[h]
	\begin{center}
		\includegraphics[width=4.2cm]{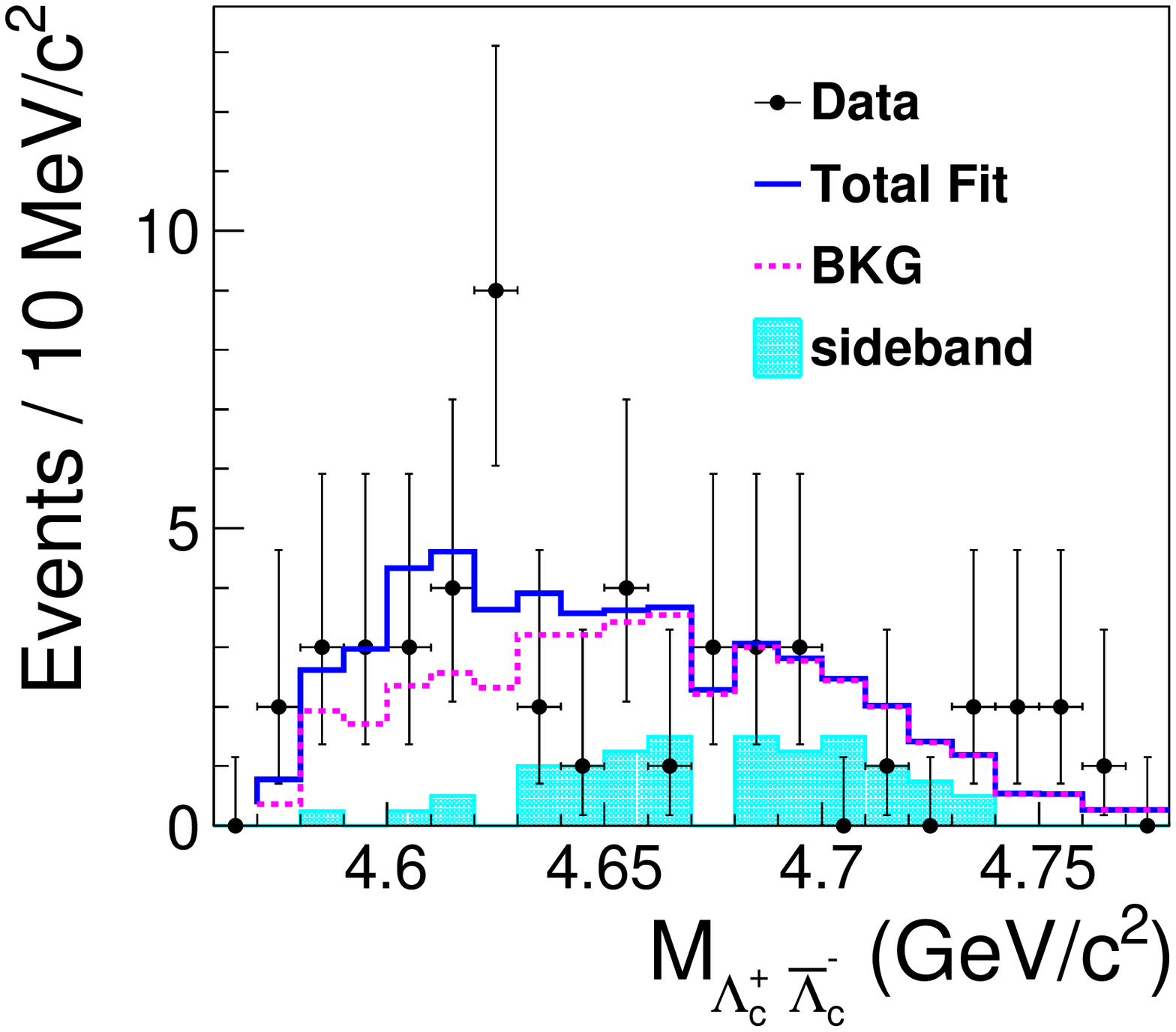}
		\includegraphics[width=4.2cm]{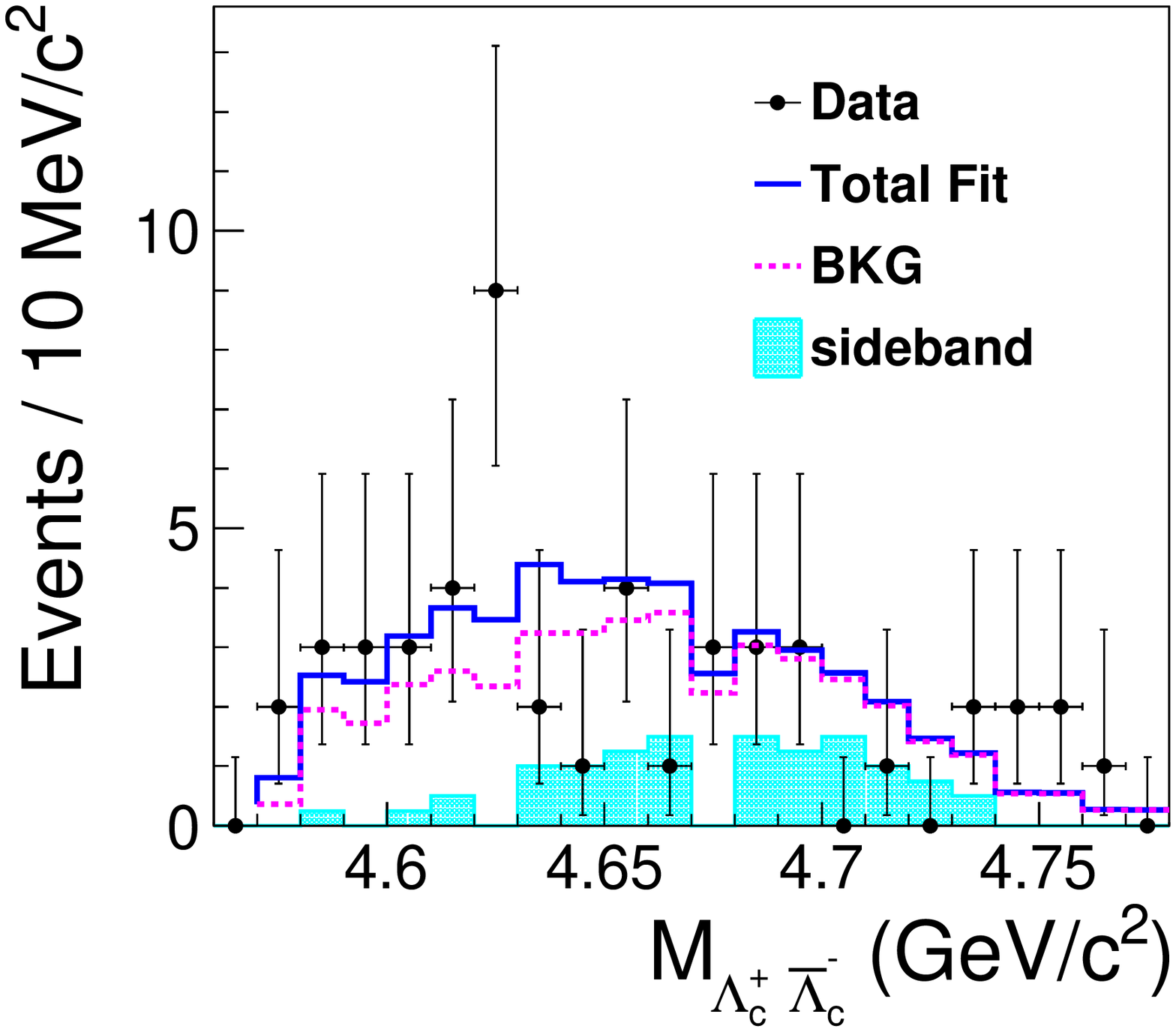}
		\put(-210,80){\bf (a)}
		\put(-90,80){\bf (b)}
		\caption{\label{Yplot} The $\Lambda_{c}^{+}\bar{\Lambda}_{c}^{-}$ invariant mass spectra in data with (a) $Y_{\eta}$ and (b) Y(4660) signals included in the fits. The solid blue lines are the best fits and the dotted red lines represent the backgrounds. The shaded cyan histograms are from the normalized $\Lambda_{c}^{+}$ and $\bar{\Lambda}_{c}^{-}$ mass sidebands.
		}
	\end{center}
\end{figure}

The $M_{\Lambda_{c}^{+}\bar{\Lambda}_{c}^{-}}$ spectrum is shown in Fig.~\ref{Yplot}, where the shaded cyan histogram is from the normalized $\Lambda_{c}^{+}$ and $\bar{\Lambda}_{c}^{-}$ mass sidebands. No evident signals of $Y_{\eta}$ or $Y(4660)$ can be seen. An unbinned extended  maximum likelihood fit is applied to the $\Lambda_{c}^{+}\bar{\Lambda}_{c}^{-}$ mass spectrum to extract the signal yields of the $Y_{\eta}$ and $Y(4660)$ in $B$ decays separately. In the fit, the signal shape of the $Y_{\eta}$ or $Y(4660)$ is obtained from MC simulation directly, with the input parameters $M_{Y_{\eta}} = 4616$ MeV/$c^{2}$ and $\Gamma_{Y_{\eta}} = 30$ MeV for $Y_{\eta}$~\cite{4616b}, and $M_{Y(4660)} = 4643$ MeV/$c^{2}$ and $\Gamma_{Y(4660)} = 72$ MeV for $Y(4660)$~\cite{PDG}. The background is described by the sum of the phase space shape, the normalized $\Lambda_{c}^{+}$ and $\bar{\Lambda}_{c}^{-}$ mass sidebands, and reflection shape of the charged $\Xi_c(2930)^+$ which has been obtained from MC simulation with the number of events fixed to that obtained from the fit to the $M_{K_{S}^{0} \Lambda_{c}^+}$ distribution. The fit results are shown in Figs.~\ref{Yplot}(a) and (b) for the $Y_{\eta}$ and $Y(4660)$, respectively. From the fits, we obtain $(10.4\pm 5.6)$ $Y_{\eta}$ and $(10.0 \pm 6.7)$ $Y(4660)$ signal events each with a signal statistical significance of $2.0 \sigma$ and $1.6 \sigma$ ($n.d.f. = 1$), respectively.

As the statistical signal significance of each $Y$ state is less than $3\sigma$, 90\% C.L. Bayesian upper limits on $\BR(\bar{B}^{0} \to \bar{K}^{0} Y)\BR( Y \to \Lambda_{c}^{+} \bar{\Lambda}_{c}^{-})$ are determined to be $2.2 \times 10^{-4}$ and $2.3 \times 10^{-4}$ for $Y=Y_\eta$ and $Y(4660)$, respectively, by solving the equation $\int_{0}^{\BR^{\rm up}}\mathcal{L}(\BR)d\BR/\int_0^{+\infty}\mathcal{L}(\BR) d\BR = 0.9$, where $\BR = \frac{n_{Y}}{\varepsilon_{\rm all}^{Y} N_{B^{0}\bar{B}^{0}}\BR(\Lambda_{c}^{+} \to p K^{-} \pi^{+})^{2}}$ is the assumed product branching fraction; $\mathcal{L}(\BR)$ is the corresponding maximized likelihood of the data; $n_{Y}$ is the number of $Y$ signal events; and $\varepsilon_{\rm all}^{Y} = \sum \varepsilon_{i}^{Y} \times \Gamma_{i}/\Gamma(p K^{-} \pi^{+})$ ($\varepsilon_{i}^{Y}$ is the detection efficiency from MC simulation for mode $i$).
To take the systematic uncertainty into account, the above likelihood is convolved with a Gaussian function whose width equals to the total systematic uncertainty discussed below.

The systematic uncertainties in the branching fraction measurements are listed below. The detection efficiency relevant (DER) uncertainties include those for tracking efficiency (0.35\%/track), particle identification efficiency (1.0\%/kaon, 0.9\%/pion, 3.7\%/proton and 3.4\%/anti-proton), as well as $\Lambda$ (3.0\%) and $K_{S}^{0}$ (2.3\%) selection efficiencies. Assuming all the above systematic uncertainty sources are independent, the DER uncertainties are summed in quadrature for each decay mode, yielding 5.8--8.6\%, depending on the mode. For the four branching fraction measurements, the final DER uncertainties are summed in quadrature over the three $\Lambda_c$ decay modes using weight factors equal to the product of the total efficiency and the $\Lambda_c$ partial decay width. Systematic uncertainties associated with the fitting procedure are estimated by a changing the order of the background polynomial, changing the range of the fit, and by enlarging the mass resolution by 10\% for all the fits; (b) adding the possible contributions from
charged $\Xi_c(2815)$ and $\Xi_c(2970)$ states in the fit to $M_{K_{S}^{0} \Lambda_{c}}$ spectrum; (c) changing the values of the masses and widths of the $Y_{\eta}$ and $Y(4660)$ by $\pm 1\sigma$ and changing the fitted number of $\Xi_c(3920)$ by 1$\sigma$ in the fit to $M_{\Lambda_{c}^{+} \bar{\Lambda}_{c}^{-}}$ spectrum. The deviations from nominal fit results are taken
as systematic uncertainties. Uncertainties for $\BR(\Lambda_c^+ \to p K^- \pi^+)$ and $\Gamma_{i}/\Gamma(p K^{-} \pi^{+})$ are taken from Ref.~\cite{PDG}. The final uncertainties on the $\Lambda_c$ partial decay widths are summed in quadrature over the three modes weighted by the detection efficiency. The world average of $\BR(\Upsilon(4S)\to B^0 \bar{B}^0)$ is $(48.6\pm0.6)\%$~\cite{PDG}, which corresponds to a systematic uncertainty of 1.23\%. The systematic uncertainty on $N_{\Upsilon(4S)}$ is 1.37\%. The total systematic uncertainties are found by adding the uncertainties from all sources in quadrature, and they are listed in Table~\ref{tab:err2}.

\begin{table}[htbp]
	\begin{threeparttable}
		\caption{\label{tab:err2}  Relative systematic
			uncertainties (\%) in the branching fraction measurements. Here,
			$\BR_1\equiv \BR (\bar{B}^{0} \to \bar{K}^{0} \Lambda_{c}^{+} \bar{\Lambda}_{c}^{-})$, $\BR_2 \equiv \BR(\bar{B}^{0}
			\to \Xi_{c}(2930)^{+} \bar{\Lambda}_{c}^{-})\BR(\Xi_{c}(2930)^{+} \to \bar{K}^0 \Lambda_{c}^+)$, $\BR_3 \equiv
			\BR(\bar{B}^0 \to \bar{K}^0 Y_{\eta})\BR( Y_{\eta} \to \Lambda_{c}^{+} \bar{\Lambda}_{c}^{-} )$, and $\BR_{4} \equiv
			\BR(\bar{B}^0 \to \bar{K}^0 Y(4660))\BR( Y(4660) \to \Lambda_{c}^{+} \bar{\Lambda}_{c}^{-} )$.
		}
		\begin{tabular}{c||c|c|c|c|cp{3cm}}
			\hline\hline
			Branching fraction & DER &  Fit  &  \tabincell{c}{$\Lambda_{c}$\\ decays} & \tabincell{c}{ $N_{B^{0}\bar{B}^{0}}$}  & \tabincell{c}{Sum} \\
			\hline
			$\BR_1$         & 5.28  & 4.20 & 10.5  & 1.82 & 12.6  \\
			$\BR_2$         & 5.31  & 6.10 & 10.5  & 1.82 & 13.4  \\
			$\BR_3$         & 5.28  & 10.2 & 10.5  & 1.82 & 15.7  \\
			$\BR_4$         & 5.27  & 11.6 & 10.5  & 1.82 & 13.3 \\
			\hline\hline
		\end{tabular}
	\end{threeparttable}
\end{table}


The sources of systematic uncertainties of charged $\Xi_{c}(2930)^+$ mass and width measurements are calculated with the following method. Half of the correction due to the input and output difference on the charged $\Xi_{c}(2930)^+$ mass determined from MC simulation is conservatively taken as a systematic uncertainty. By enlarging the mass resolution by 10\%, the difference in the measured $\Xi_{c}(2930)^+$ width is 0.9 MeV, and this is taken as a systematic uncertainty. By changing the background shape, the differences of 0.5 MeV/$c^2$ and 1.3 MeV in the measured charged $\Xi_{c}(2930)^+$ mass and width, respectively, are taken as systematic uncertainties.

The signal-parametrization systematic uncertainty is estimated by replacing the constant total width with a mass-dependent width of $\Gamma_{t} = \Gamma^{0}_{t}\times\Phi(M_{K^{0}_{S}\Lambda_{c}^+})/\Phi(M_{\Xi_c(2930)^+})$, where $\Gamma_{t}^{0}$ is the width of the resonance, $\Phi(M_{K^{0}_{S}\Lambda_c^+}) = P/M_{K^{0}_{S}\Lambda_c^+}$ is the phase space factor for an S-wave two-body system ($P$ is the $K^{0}_{S}$ momentum in the $K^{0}_{S} \Lambda_c^+$ CMS) and $M_{\Xi_c(2930)^+}$ is the $K^{0}_{S} \Lambda_c^+$ invariant mass fixed at the charged $\Xi_{c}(2930)^+$ nominal mass. Due to the limited statistic, we generate $K_{S}^{0} \Lambda_{c}^{+}$ mass spectrum according to the fitted $\Xi_{c}(2930)^{+}$ shape with 200 times of events than the fitted signal yield. By fitting this mass spectrum with mass-dependent RBW function, the difference in the measured
$\Xi_{c}(2930)^+$ mass is negligible and the difference in the width is 1.9 MeV which is taken as the systematic uncertainty.
Assuming all the sources are independent, we add them in quadrature to obtain the total systematic uncertainties on the charged $\Xi_{c}(2930)^+$ mass and width of $1.5$ MeV/$c^2$ and $2.5$ MeV, respectively.


In summary, using $(772 \pm 11) \times 10^{6}~B\bar{B}$ pairs, we perform an updated analysis of $\bar{B}^{0} \to \bar{K}^{0} \Lambda_{c}^{+} \bar{\Lambda}_{c}^{-}$. There is $4.1\sigma$ evidence of the charged charmed baryon state $\Xi_{c}(2930)^+$ in the $K^0_{S} \Lambda_{c}^+$ mass spectrum. The measured mass and width are $M_{\Xi_{c}(2930)^{+}} =[2942.3\pm 4.4  (\rm stat.) \pm 1.5(\rm syst.)]$~MeV/$c^{2}$ and $\Gamma_{\Xi_{c}(2930)^{+}} = [14.8 \pm 8.8(\rm stat.) \pm 2.5(\rm syst.)]$ MeV. The mass and width difference between neutral and charged $\Xi_{c}(2930)$ is $\Delta m = [-13.4 \pm 5.3(\rm stat.) ^{+1.7}_{-12.1}(\rm syst.)]$~MeV/$c^{2}$ and $\Delta \Gamma = [4.7 \pm 12.2(\rm stat.)^{+6.4}_{-8.3}(\rm syst.)]$ MeV, respectively. The branching fraction is $\BR(\bar{B}^{0} \to \bar{K}^{0} \Lambda_{c}^{+}\bar{\Lambda}_{c}^{-})=[3.99 \pm 0.76(\rm stat.) \pm 0.51(\rm syst.)]\times 10^{-4} $, which is consistent with the world average value of $(4.3 \pm 2.2) \times 10^{-4}$~\cite{PDG} but with much improved precision. We measure the product branching fraction $\BR(\bar{B}^{0} \to \Xi_{c}(2930)^{+} \bar{\Lambda}_{c}^{-})\BR(\Xi_{c}(2930)^{+} \to \bar{K}^{0} \Lambda_{c}^{+})=[2.37 \pm 0.51 (\rm stat.)\pm 0.31(\rm syst.)] \times 10^{-4}$. Due to the limited statistics, we are not able to perform an angular analysis to determine the spin-parity of the $\Xi_{c}(2930)^{+}$, and cannot identify its quark configuration for which there are many theoretical possibilities.	We expect that a spin-parity analysis will be possible with the much larger data sample which will be collected with the Belle II detector. There are no significant signals seen in the $\Lambda_{c}^{+} \bar{\Lambda}_{c}^{-}$ mass spectrum. We place 90\% C.L. upper limits for the $Y(4660)$ and its theoretically predicted spin partner $Y_{\eta}$ of $\BR(\bar{B}^0 \to \bar{K}^0 Y(4660))\BR(Y(4660) \to \Lambda_{c}^{+} \bar{\Lambda}_{c}^{-}) < 2.3 \times 10^{-4}$ and $\BR(\bar{B}^0 \to \bar{K}^0 Y_{\eta})\BR( Y_{\eta} \to \Lambda_{c}^{+} \bar{\Lambda}_{c}^{-}) < 2.2 \times 10^{-4}$~\cite{BBB2}.


%


\begin{thebibliography}{99}
\bibitem{charge-conjugate} Inclusion of charge conjugate states is implicit unless otherwise stated.
\bibitem{BBB2} Considering the possible change of $\BR(\Lambda_{c}^{+} \to p K^{-} \pi^{+})$ in the future, we provide $\BR(\bar{B}^{0} \to \bar{K}^{0} \Lambda_{c}^{+} \bar{\Lambda}_{c}^{-})\BR(\Lambda_{c}^{+} \to p K^{-} \pi^{+})^{2} = (1.55 \pm 0.29 \pm 0.11)\times 10^{-6}$ and $\BR(\bar{B}^{0} \to \Xi_{c}(2930)^{+} \bar{\Lambda}_{c}^{-})\BR(\Xi_{c}(2930)^{+}\to \bar{K}^0 \Lambda_{c}^{+})\BR(\Lambda_{c}^{+} \to p K^{-} \pi^{+})^{2} = (9.18 \pm 2.0 \pm 0.8) \times 10^{-7}$, where the first errors are statistical and the second systematic with the uncertainty on $\BR(\Lambda_{c}^{+} \to p K^{-} \pi^{+})$ omitted. The 90\% C.L. upper limits on $\BR(\bar{B}^0 \to \bar{K}^0 Y(4660))\BR(Y(4660) \to \Lambda_{c}^{+} \bar{\Lambda}_{c}^{-})\BR(\Lambda_{c}^{+} \to p K^{-}\pi^{+})^{2}$ and $\BR(\bar{B}^0 \to \bar{K}^0 Y_{\eta})\BR( Y_{\eta} \to \Lambda_{c}^{+} \bar{\Lambda}_{c}^{-})  \BR(\Lambda_{c}^{+} \to p K^{-} \pi^{+})^{2}$ are $8.9 \times 10^{-7}$ and $8.5 \times 10^{-7}$, respectively.
\end{thebibliography}
\end{document}